\documentclass[obeyspaces,
  journal=pasa,
  manuscript=research-paper, 
  year=202X,
  volume=YY,
]{cup-journal}

\usepackage{lineno}
\usepackage{amsmath}
\usepackage[nopatch]{microtype}
\usepackage{booktabs}
\usepackage{multirow}
\usepackage{gensymb}
\usepackage{url}
\usepackage{amssymb,microtype,siunitx,booktabs}
\usepackage[colorlinks=true,linkcolor=blue]{hyperref}
\hypersetup{citecolor=blue, urlcolor=blue}

\newcommand\blfootnote[1]{%
  \begingroup
  \renewcommand\thefootnote{}\footnote{#1}%
  \addtocounter{footnote}{-1}%
  \endgroup
}

\def\II{\hbox{\sc ii}}
\newcommand{\farcm}{\mbox{\ensuremath{.\mkern-4mu^\prime}}}
\newcommand{\arcsec}{\mbox{\ensuremath{^{\prime\prime}}}}
\newcommand{\farcs}{\mbox{\ensuremath{.\!\!^{\prime\prime}}}}
\newcommand{\fdg}{\mbox{\ensuremath{.\!\!^\circ}}}
\newcommand{\kms}{\,km\,s$^{-1}$}
\newcommand{\mujybm}{\,$\mu$Jy\,beam$^{-1}$}

\def\HII{\hbox{H{\sc ii}}}

\usepackage[nolist]{acronym}
\begin{acronym}[AWGN]
\acro{ASKAP}{Australian Square Kilometre Array Pathfinder}
\acro{EMU}{Evolutionary Map of the Universe}
\acro{DR3}{Data Release 3}
\acro{rms}{root mean squared}
\acro{WR}{Wolf-Rayet}
\acro{PM}{proper motion}
\acro{ACIS-I}{Advanced CCD Imaging Spectrometer using the wide field chip}
\acro{CALDB}{Chandra Calibration Database}
\acro{CARTA}{Cube Analysis and Rendering Tool for Astronomy}
\acro{CASDA}{CSIRO \ac{ASKAP} Science Data Archive}
\acro{CIAO}{Chandra Interactive Analysis of Observations software}
\acro{CDA}{Chandra Data Archive}
\acro{FoV}{field of view}
\acro{ISM}{interstellar medium}
\acro{MGPS}{MeerKAT Galactic Plane Survey}
\acro{MGPS-2}{Molonglo Galactic Plane Survey epoch-2}
\acro{NASA}{National Aeronautics and Space Administration}
\acro{NS}{neutron star}
\acro{ObsID}{observation identification}
\acro{PA}{position angle}
\acro{PAF}{phased array feed}
\acro{RFI}{radio-frequency interference}
\acro{rms}{root-mean-square}
\acro{SB}{Scheduling block}
\acro{UWL}{Ultra-Wideband Low}
\acro{WISE}{Wide-Field Infrared Survey Explorer}
\acro{2MASS}{Two Micron All Sky Survey}
\acro{2MASX}{Two Micron All Sky Survey Extended Source Catalogue}
\acro{30Dor}[30 Dor]{30~Doradus}
\acro{AGN}{active galactic nuclei}
\acro{ATCA}{Australia Telescope Compact Array}
\acro{ATESP}{Australia Telescope ESO Slice Project}
\acro{ATNF}{Australia Telescope National Facility}
\acro{ATOA}{Australia Telescope Online Archive}
\acro{AT20G}{Australia Telescope 20 GHz Survey}
\acro{ASKAP}{Australian Square Kilometre Array Pathfinder}
\acro{ARC}{Australian Research Council}
\acro{BETA}{Boolardy Engineering Test Array}
\acro{BL}[BL Lac]{BL Lacertae Objects}
\acro{CASS}{CSIRO Astronomy and Space Science}
\acro{CABB}{Compact Array Broadband Back-end}
\acro{CHII}[{\sc CHii}]{Compact \textsc{Hii}}
\acro{CMB}{Cosmic Microwave Background}
\acro{CSIRO}{Australian Commonwealth Scientific and Industrial Research Organisation}
\acro{CSS}{Compact Steep Spectrum}
\acro{DS9}[\textsc{DS9}]{\textsc{SAOImage DS9}}
\acro{DSA}{diffusive shock acceleration}
\acro{DSS}{Digital Sky Survey}
\acro{EM}{Electromagnetic}
\acro{EMU}{Evolutionary Map of the Universe}
\acro{eROSITA}{extended R\"Ontgen Survey with an Imaging Telescope Array}
\acro{ev}[eV]{electronvolt\acroextra{: 1 eV $\approx 1.6 \times 10^{-19}$ J}}
\acro{FITS}[\textsc{Fits}]{Flexible Image Transport System}
\acro{FRB}{Fast Radio Bursts}
\acro{FSRQ}{Flat Spectrum Radio Quasars}
\acro{FWHM}{Full Width at Half-Maximum}
\acro{GLEAM}{GaLactic Extragalactic All-sky MWA}
\acro{GPS}{Gigahertz Peak Spectrum}
\acro{HEMT}{High Electron Mobility Transistor}
\acro{H.E.S.S.}{High Energy Spectroscopic System}
\acro{HFP}[HFP]{High-Frequency Peaker}
\acro{HPBW}{Half Power Beam Width} 
\acro{HzRGs}{High Redshift Radio Galaxies}
\acro{pHFP}[pHFP]{Potential High Frequency Peaker}
\acro{HI}[H{\sc i}]{Neutral Atomic Hydrogen} 
\acro{HST}{\textit{Hubble Space Telescope}}
\acro{ICM}{intracluster medium}
\acro{IAU}{International Astronomical Union}
\acro{IFRSs}{Infrared Faint Radio Sources}
\acro{ISM}{Interstellar Medium}
\acro{IFRS}{Infrared Faint Radio Source}
\acro{IR}{Infrared}
\acro{JY}[Jy]{Jansky\acroextra{, 1 Jy = $10^{-26} \times \mathrm{W~ m}^{-2}~\mathrm{Hz}^{-1}$}} 
\acro{LBV}{Luminous Blue Variable}
\acro{LLS}{Largest Linear Size}
\acro{LFAA}{Low-Frequency Aperture Array}
\acro{LMC}{Large Magellanic Cloud}
\acro{LOFAR}{Low-Frequency Array}
\acro{LSO}[LSOs]{Large Scale Objects}
\acro{MACHO}{Massive Astrophysical Compact Halo Objects}
\acro{MC}[MCs]{Magellanic Clouds}
\acro{mc}[MC]{Magellanic Cloud}    
\acro{MCELS}{Magellanic Cloud Emission Line Survey}
\acro{MW}{Milky Way}
\acro{MIRIAD}[\textsc{Miriad}]{Multichannel Image Reconstruction, Image Analysis and Display}
\acro{MIT}{Massachusetts Institute of Technology}
\acro{MOST}{Molonglo Observatory Synthesis Telescope}
\acro{MQS}{Magellanic Quasars Survey}
\acro{MRC}{Molonglo Reference Catalogue of Radio Sources}
\acro{MWA}{Murchison Widefield Array}
\acro{NRAO}{National Radio Astronomy Observatory}
\acro{NVSS}{NRAO VLA Sky Survey}
\acro{OPAL}{Online Proposal Applications \& Links}
\acrodefplural{ORC}{Odd Radio Circles}
\acro{ORC}{Odd Radio Circle}
\acro{OVV}{Optically Violent Variable Quasars}            
\acro{PAF}{Phased Array Feed}
\acro{pc}{parsec\acroextra{: 1 pc $\simeq 3.09 \times 10^{16}$ m}}
\acro{PMN}{Parkes-MIT-NRAO}
\acro{PNe}{Planetary Nebulae}
\acro{PWN}{Pulsar Wind Nebulae}
\acro{QSO}{Quasi-Stellar Object}
\acro{RA}{Right Ascension}
\acro{RFI}{Radio-Frequency Interference}
\acro{RMS}[rms]{Root Mean Squared}
\acro{SARAO}{South African Radio Astronomy Observatory}
\acro{SDSS}{Sloan Digital Sky Survey}
\acro{SED}{spectral energy distribution}
\acro{SI}[$\alpha$]{Spectral Index\acroextra{, $S \propto \nu^\alpha$}}
\acro{SKA}{Square Kilometre Array}
\acro{SMB}{Super Massive Blackholes}
\acro{SMC}{Small Magellanic Cloud}
\acro{SN}{Supernova}
\acro{SUMSS}{Sydney University Molonglo Sky Survey}
\acro{TOPCAT}[\textsc{Topcat}]{Tool for OPerations on Catalogues And Tables}
\acro{USS}{Ultra Steep Spectrum}
\acro{YSOs}{young stellar objects}
\acro{WBAC}{Wide-Band Analogue Correlator}
\acro{WIFES}[WiFeS]{Wide-Field Spectrograph}
\acro{WISE}{Wide-Field Infrared Survey Explorer}
\acro{VLA}{Very Large Array} 
\acro{VLBI}{Very Long Baseline Interferometry} 
\acro{VLSR}[\textbf{$v_{lsr}$}]{Velocity in the Line of Sight}
\acro{SMASH}{Survey of the MAgellanic Stellar History}
\acro{SNR}{Supernova Remnant}
\acrodefplural{SNR}{Supernova Remnants}
\acro{SD}{single-degenerate}
\acro{SUMSS}{Sydney University Molonglo Sky Survey}
\acro{SMBH}{Super Massive Black Hole}
\acro{RSG}{Red Super Giant}
\end{acronym}

\title{Evolutionary Map of the Universe: Detection and Analysis of the Shell Surrounding the Runaway Wolf-Rayet Star WR16}

\author{A. C. Bradley}
\affiliation{Western Sydney University, Locked Bag 1797, Penrith South DC, NSW 2751, Australia}
\email[A. C. Bradley]{20295208@student.westernsydney.edu.au}

\author{M. D. Filipovi{\' c}}
\affiliation{Western Sydney University, Locked Bag 1797, Penrith South DC, NSW 2751, Australia}

\author{Z. J. Smeaton}
\affiliation{Western Sydney University, Locked Bag 1797, Penrith South DC, NSW 2751, Australia}

\author{H. Sano}
\affiliation{Faculty of Engineering, Gifu University, 1-1 Yanagido, Gifu 501-1193, Japan}
\alsoaffiliation{National Astronomical Observatory of Japan, Mitaka, Tokyo 181-8588, Japan}

\author{Y. Fukui}
\affiliation{Faculty of Engineering, Gifu University, 1-1 Yanagido, Gifu 501-1193, Japan}
\alsoaffiliation{Department of Physics, Nagoya University, Furo-cho, Chikusa-ku, Nagoya 464-8601, Japan}

\author{C. Bordiu}
\affiliation{INAF\,$-$\,Osservatorio Astrofisico di Catania, Via S. Sofia 78, I-95123, Catania, Italy}

\author{S. Cichowolski}
\affiliation{Instituto de Astronomía y Física del Espacio (UBA, CONICET), CC 67, Suc. 28, 1428 Buenos Aires, Argentina}

\author{N. F. H. Tothill}
\affiliation{Western Sydney University, Locked Bag 1797, Penrith South DC, NSW 2751, Australia}

\author{R. Z. E. Alsaberi}
\affiliation{Faculty of Engineering, Gifu University, 1-1 Yanagido, Gifu 501-1193, Japan}
\alsoaffiliation{Western Sydney University, Locked Bag 1797, Penrith South DC, NSW 2751, Australia}

\author{F. Bufano}
\affiliation{INAF\,$-$\,Osservatorio Astrofisico di Catania, Via S. Sofia 78, I-95123, Catania, Italy}

\author{S. Dai}
\affiliation{Australia Telescope National Facility, CSIRO, Space and Astronomy, PO Box 76, Epping, NSW 1710, Australia}

\author{Y. A. Gordon}
\affiliation{Department of Physics, University of Wisconsin-Madison, 1150 University Avenue, Madison, WI 53706, USA}

\author{A. M. Hopkins}
\affiliation{School of Mathematical and Physical Sciences, 12 Wally’s Walk, Macquarie University, NSW 2109, Australia}

\author{T. H. Jarrett$^\dag$}
\affiliation{Department of Astronomy, University of Cape Town, Private Bag X3, Rondebosch 7701, South Africa}
\alsoaffiliation{Western Sydney University, Locked Bag 1797, Penrith South DC, NSW 2751, Australia}

\author{B. S. Koribalski}
\affiliation{Australia Telescope National Facility, CSIRO, Space and Astronomy, PO Box 76, Epping, NSW 1710, Australia}
\alsoaffiliation{Western Sydney University, Locked Bag 1797, Penrith South DC, NSW 2751, Australia}

\author{S. Lazarevi\'c}
\affiliation{Western Sydney University, Locked Bag 1797, Penrith South DC, NSW 2751, Australia}
\alsoaffiliation{Australia Telescope National Facility, CSIRO, Space and Astronomy, PO Box 76, Epping, NSW 1710, Australia}
\alsoaffiliation{Astronomical Observatory, Volgina 7, 11060 Belgrade, Serbia}

\author{C. J. Riseley}
\affiliation{Astronomisches Institut der Ruhr-Universit\"{a}t Bochum (AIRUB), Universit\"{a}tsstra{\ss}e 150, 44801 Bochum, Germany}

\author{G. Rowell}
\affiliation{School of Physical Sciences, The University of Adelaide, Adelaide 5005, Australia}

\author{M. Sasaki}
\affiliation{Dr Karl Remeis Observatory, Erlangen Centre for Astroparticle Physics, Friedrich-Alexander-Universit\"{a}t Erlangen-N\"{u}rnberg, Sternwartstra{\ss}e 7, 96049 Bamberg, Germany}

\author{D. Uro\v{s}evi\'c}
\affiliation{Department of Astronomy, Faculty of Mathematics, University of Belgrade, Studentski trg 16, 11000 Belgrade, Serbia}

\author{T. Vernstrom}
\affiliation{CSIRO Space and Astronomy, PO Box 1130, Bentley, WA 6102, Australia}
\alsoaffiliation{ICRAR, The University of Western Australia, 35 Stirling Hw, 6009 Crawley, Australia}

\received {dd Mmm YYYY}
\revised  {dd Mmm YYYY}
\accepted {dd Mmm YYYY}
\published{dd Mmm YYYY}


\begin{document}

\begin{abstract}
We present the first radio--continuum detection of the circumstellar shell around the well-known WN8 type Wolf-Rayet star WR16 at 943.5\,MHz using the \ac{ASKAP} \ac{EMU} survey. At this frequency, the shell has a measured flux density of 72.2$\pm$7.2\,mJy. Using previous \ac{ATCA} measurements at 2.4, 4.8, and 8.64~GHz, as well as the \ac{EMU} observations of the star itself, we determine a spectral index of $\alpha\,=\,+0.74\pm0.02$, indicating thermal emission. We propose that the shell and star both exhibit thermal emission, supported by the its appearance in near-infrared and H$\alpha$ observations. The latest \textit{Gaia} parallax is used to determine a distance of 2.28$\pm$0.09\,kpc. This star is well-known for its surrounding circular nebulosity, and using the distance and an angular diameter of 8\farcm42, we determine the shell size to be 5.57$\pm$0.22~pc. We use the \textit{Gaia} \acp{PM} of WR16 to determine peculiar velocities of the star as $V_{\alpha}(pec) =$ --45.3$\pm$5.4\,\kms\ and $V_{\delta}(pec) =$ 22.8$\pm$4.7\,\kms, which indicates that the star is moving in a north-west direction, and translates to a peculiar tangential velocity to be 50.7$\pm$6.9\,\kms. We also use these \acp{PM} to determine the shell's origin, estimate an age of $\sim 9500\pm 1300$\,yr, and determine its average expansion velocity to be $280\pm40$\,\kms. This average expansion velocity suggests that the previous transitional phase is a \ac{LBV} phase, rather than a \ac{RSG} phase. We also use the measured flux at 943.5~MHz to determine a mass-loss rate of $1.753\times 10^{-5}~M_\odot~$yr$^{-1}$, and use this to determine a lower-limit on ionising photons of $N_{UV} > 1.406\times 10^{47}~$s$^{-1}$.
\end{abstract}

\acresetall

\section{Introduction}
\label{Section:Introduction}

\blfootnote{$^\dag$Deceased on 3$^{\rm{rd}}$ July 2024}\ac{WR} stars are massive late stage stars with broad emission lines \citep{2007ARA&A..45..177C}. They are categorised by their dominating elemental emission: carbon (WC), nitrogen (WN), and oxygen (WO). \ac{WR} stars follow an evolutionary cycle that is identifiable by its previous mass--loss history, where mass-loss outbursts are inconsistent unlike O-stars, and caused by internal pulsations \citep{1994A&A...290..819L}. The previous mass--loss history of these stars is easily identifiable by outbursts of stellar material surrounding them \citep{2000A&A...360..227N}.


WR16 is a \ac{WR} star of the WN8 type \citep{2009AJ....138..402S} at RA~(J2000)~=~ 09:54:52.91, Dec~(J2000)~=~--57:43:38.30~\citep{2015A&A...578A..66T}, which is well known for its ring-like nebula \citep{2020MNRAS.495..417C}. While the inner nebula is circular, it is proposed that two more arc-like features outside of this inner nebula share the same origin \citep[Figure~\ref{fig:WR16/WR40};][]{1995AJ....109.1839M}. It is likely that winds from the previous O-type star main sequence would have swept the surrounding ISM over several Myrs, creating  a wind-blown shell \citep{1994A&A...290..819L}. Then, after the main sequence, the star would have entered a transitional phase, suffering a number of mass-loss events/outbursts \citep{2007ARA&A..45..177C}. The ejecta from these outbursts would have expanded freely within the cavity carved by the main sequence wind. Finally, when the star became a \ac{WR}, the \ac{WR} wind pushed and compressed the ejecta, allowing its most recent outburst to expand symmetrically \citep{1995AJ....109.1839M}. 

The \ac{ASKAP} \citep{2021PASA...38....9H} \ac{EMU} \citep{Norris2011,Norris2021,2025arXiv250508271H} survey is mapping the entire southern sky at 943.5\,MHz and is currently $\sim$25$\%$ complete. The \ac{ASKAP} telescope has good surface brightness sensitivity that allows us to see characteristics of objects not detected before at this frequency. Previously, extended sources with a similar low-surface brightness to WR16's shell have been characterised such as \acp{SNR};
J0624--6948~\citep{2022MNRAS.512..265F,2025A&A...693L..15S}, G288.8--6.3~\citep[Ancora;][]{2023AJ....166..149F,2024A&A...684A.150B}, G308.7+1.4 \citep[Raspberry;][]{2024RNAAS...8..107L}, G312.6+ 2.8 \citep[Unicycle;][]{2024RNAAS...8..158S}; and a pulsar wind nebula (PWN) \citep[Potoroo;][]{2024PASA...41...32L}.


We present an analysis of the star WR16 as seen by \ac{ASKAP} during mapping for the \ac{EMU} survey. In section~\ref{Section:Data}, we discuss the observing parameters of the survey, as well as outlining other data used. In section \ref{Sub_Section:Results} we explore the results from analysing the \ac{EMU} tile, as well as exploring the movement of the shell and star, the shell's expansion, and the presence of a CO bubble. Finally, we give our summary and conclusions in section \ref{Section:Conclusion}.


\section{Data}
\label{Section:Data}

\subsection{ASKAP EMU}
\label{Sub_Section:Data_ASKAP_EMU}

The object WR16 has been seen in three \ac{ASKAP} \ac{EMU} observations. SB46953 observed the tile EMU$\_1017$-$60$ on December~12$^{th}$~2022, SB51428 observed the tile EMU$\_0954$-$55$ on July~13$^{th}$~2023, and SB53568 observed the tile EMU$\_0936$-$60$ on October~7$^{th}$~2023. SB46953 is excluded from analysis due to its proximity to the tile's edge, which can cause a drop in sensitivity.

The data were reduced using the standard \ac{ASKAP} pipeline, ASKAPSoft, using multi-frequency synthesis imaging, multi-scale cleaning, self-calibration and convolution to a common beam size \citep{2019ascl.soft12003G}. As the source was observed in two usable \ac{EMU} fields, the images were combined with the radio imaging software Miriad~\citep{miriad} using the task \textsc{imcomb}. The final image was created by combining the observations with equal weighting. We measure a number of background regions around WR16 to find the background difference in the images. We find a median rms value of 37.5\,$\mu$Jy\,beam$^{-1}$ and a mean value of 37\,$\mu$Jy\,beam$^{-1}$, compared with the previous values of median\,=\,36\,$\mu$Jy\,beam$^{-1}$ and mean\,=\,43.5\,$\mu$Jy\,beam$^{-1}$ (SB51428), and median\,=\,46.5\,$\mu$Jy\,beam$^{-1}$ and mean\,=\,50.5\,$\mu$Jy\,beam$^{-1}$ (SB53568). This background noise level is slightly higher than typical for \ac{EMU} \citep[$\sigma=20-30\,\mu$Jy\,beam$^{-1}$;][]{2025arXiv250508271H}, which can be attributed to WR16 being in an area of higher background Galactic emission ($b =$~$-$2.55). The final image is shown in Figure~\ref{fig:WR16/WR40} with a resulting sensitivity of $\sigma$=37\mujybm\ and a synthesised beam of 15\arcsec$\times$15\arcsec. 


\subsection{\textit{Gaia}}
\label{Sub_Section:Data_Gaia}

We use measurements made by the \textit{Gaia} astrometry satellite \citep{2016A&A...595A...1G} for WR16. We obtain \ac{PM} and parallax values from \ac{DR3} \citep{2023A&A...674A...1G}, and use them to determine the star's movement and distance. Discussion surrounding the use of this specific data release is found in section~\ref{Sub_Section:Measurements}.


\subsection{Nanten CO}
\label{Sub_Section:Data_Nanten}

In order to explore the distribution of molecular clouds surrounding WR16, we analysed the archival $^{12}$CO($J$~=~1--0) data taken by the NANTEN 4-m radio telescope \citep{NANTEN}. The angular resolution of the data cube was 2\farcm6. The typical noise fluctuations were $\sim$0.3 K at a velocity resolution of 0.65~km~s$^{-1}$.

\subsection{Other Data}
\label{Sub_Section:Data_Other}

\begin{figure*}[ht]
\centering
    \includegraphics[width=\linewidth]{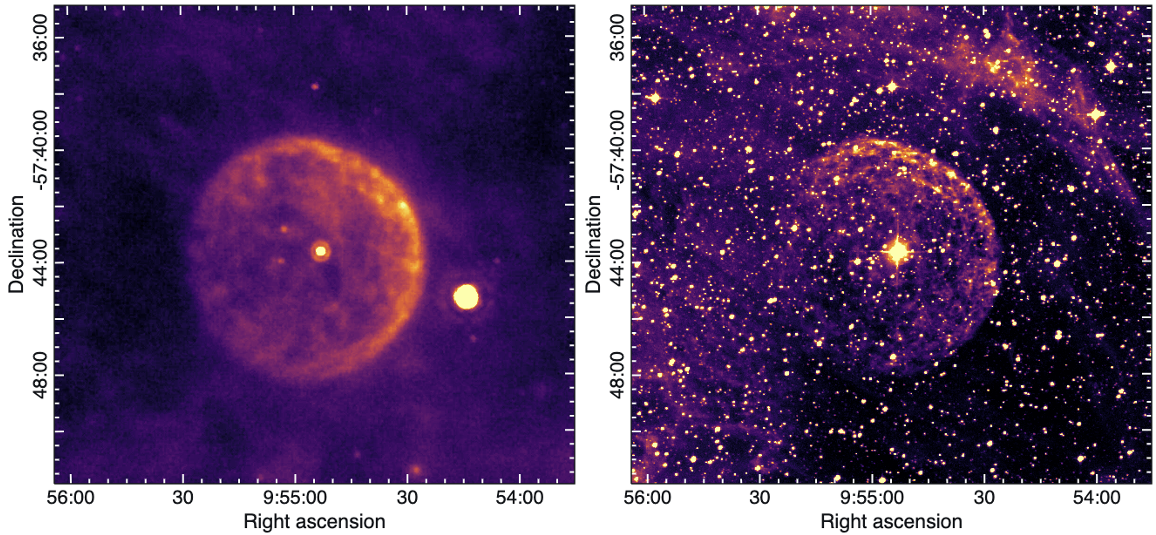}
    \caption{Detection of WR16 and its shell at 22$\mu$m from \ac{WISE} infrared (Left), and SuperCOSMOS H$\alpha$ (Right). Both sub-images are linearly scaled.}
    \label{fig:IR-Ha}
\end{figure*}

We include observations from the SuperCOSMOS \citep{2001MNRAS.326.1279H} and \ac{WISE} \citep{2010AJ....140.1868W} sky surveys to compare the nebulosity surrounding WR16 at 656.281~nm and 22\,$\mu$m (Figure~\ref{fig:IR-Ha}). 

We also use previous flux density measurements using \ac{ATCA} from \citet{1995ApJ...450..289L}, \citet{1997ApJ...481..898L}, and \citet{1999ApJ...518..890C}. These measurements are included in Table \ref{Table:Data_Fluxes}, and discussed in section \ref{Sub_Section:Measurements}.

We have searched for high-energy associations to WR16 and its shell. Fermi-LAT Data Release 4 \citep{2022ApJS..260...53A,2023arXiv230712546B} shows no corresponding Gamma-Ray emission. The SRG/eROSITA all-sky survey (eRASS) Data Release 1 \citep{2024A&A...682A..34M}, and observations taken with \textit{XMM-Newton} \citep{2001A&A...365L..18S,2001A&A...365L..27T} show no corresponding X-Ray emission.

\begin{table*}
\caption{Measured flux densities of the Wolf-Rayet star WR16 and its nebulous shell. $^a$Measurements made as part of this paper. $^b$Measurements using \ac{ATCA} observations from \citet{1999ApJ...518..890C}. $^c$Measurements using \ac{ATCA} observations from \citet{1995ApJ...450..289L}.}
\vskip.25cm
\centerline{\begin{tabular}{rccccc}
\hline
& $S_{944}$$^{a}$ & $S_{2.40}$$^{b}$ & $S_{4.80}$$^{c}$ & $S_{8.64}$$^{c}$ & $\alpha$ \\
WR16 & (mJy) & (mJy) & (mJy) & (mJy) & $\alpha\pm\Delta\alpha$ \\
\hline\hline
Star & 0.35$\pm$0.04 & 0.69$\pm$0.15 & 1.21$\pm$0.09 & 1.75$\pm$0.09 & 0.74$\pm$0.02 \\
\hline
Shell & 72.2$\pm$7.2 & -- & -- & -- & -- \\
\hline
\end{tabular}}
\label{Table:Data_Fluxes}
\end{table*}

\begin{figure*}[!ht]
\centering
    \includegraphics[width=\linewidth]{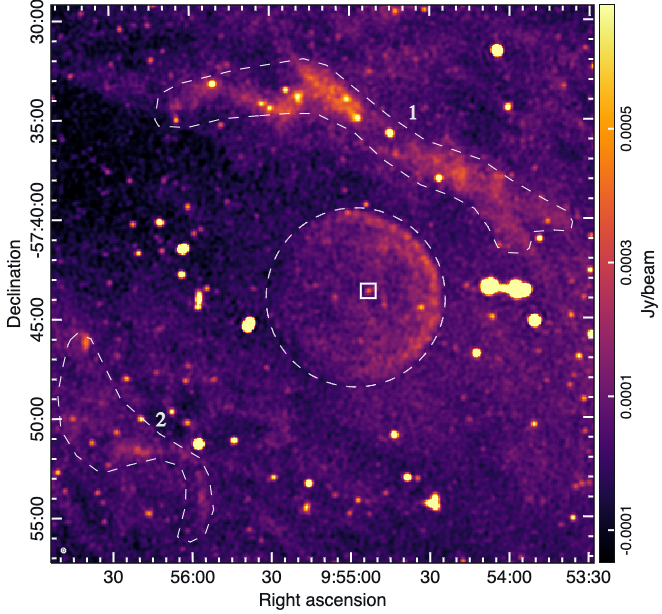}
    \caption{\ac{EMU} detection of WR16 and its inner shell at 943.5\,MHz. The image is linearly scaled and the 15$^{\prime\prime}$ beam size is presented by the small white circle in the bottom-left of the image. The white square indicates the position of WR16, the dashed circle shows the position of the `inner' circular shell. The dashed polygons labelled 1 and 2 indicate the position of `outer' shell remnants (described in \citep{1995AJ....109.1839M, 2020MNRAS.495..417C}).}
    \label{fig:WR16/WR40}
\end{figure*}


\begin{figure*}[ht]
\centering
    \includegraphics[width=\linewidth]{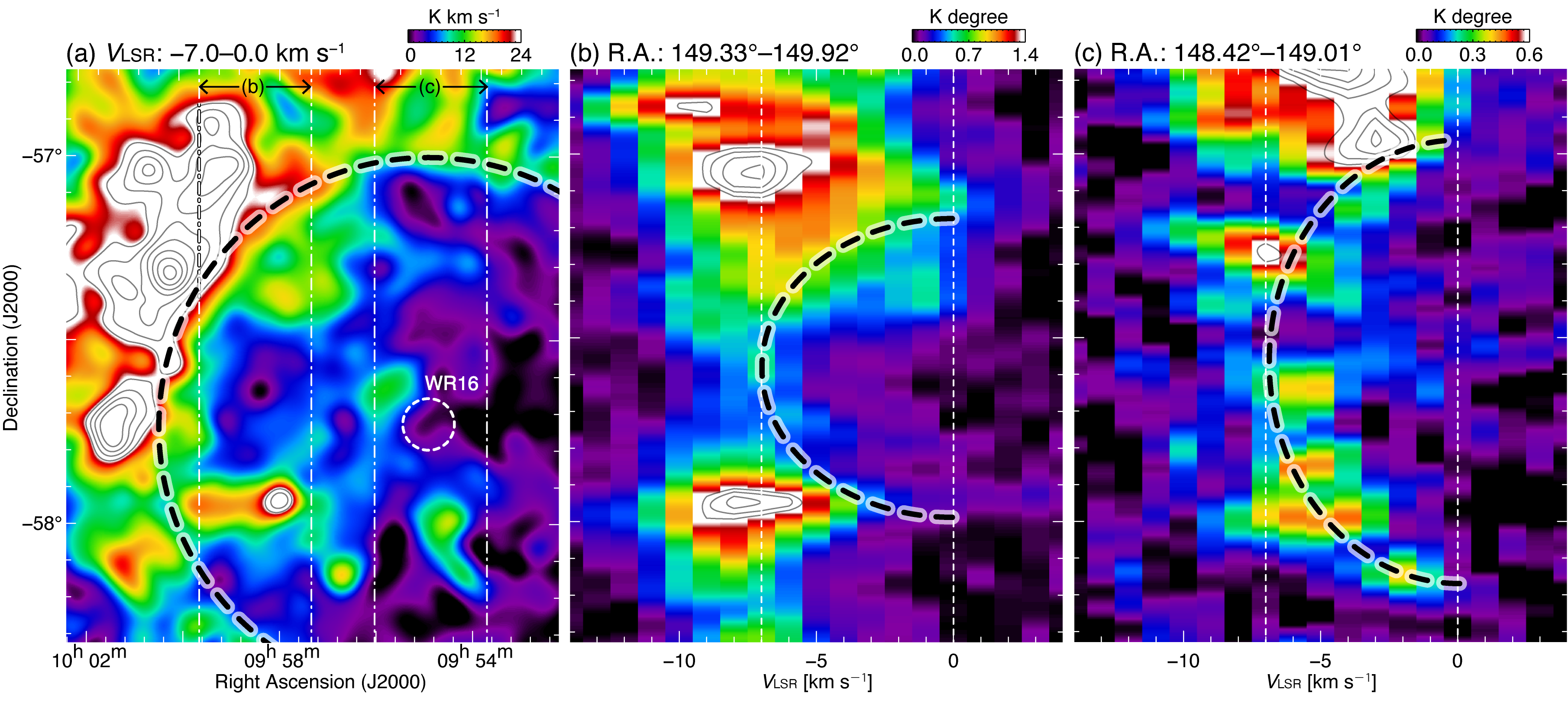}
    \caption{\textbf{(a)} -- Distribution of the NANTEN $^{12}$CO($J$~=~1--0) line emission toward WR16 \citep{NANTEN}. The integration velocity range is from $-7$ to 0~km~s$^{-1}$. The incomplete circular shell indicates a wind-blown bubble detected in CO (see section \ref{Sub_Section:CO_Map_Discussion}). The dashed circle shows the position of the inner circular shell of WR16. \textbf{(b,c)} -- Position--velocity diagrams of CO. The integration range of Right Ascension is from 149\fdg33 to 149\fdg92 for (b) and from 148\fdg42 to 149\fdg01 for (c). The dashed semicircles represent the expanding motion of CO due to the stellar wind from WR16 (see section \ref{Sub_Section:CO_Map_Discussion}).}
    \label{Figure:CO_Map}
\end{figure*}


\section{Results and Discussion}
\label{Sub_Section:Results}

\subsection{Measurements}
\label{Sub_Section:Measurements}

Using a \textit{Gaia} parallax of 0.4380$\pm$0.0168\,mas \citep{2020yCat.1350....0G}, we estimate the distance of WR16 to be 2.28$\pm$0.09\,kpc. This is in accordance with the photometric distance calculated in \citet{2001NewAR..45..135V} of 2.37\,kpc, and with a distance of 2.23$\pm$0.39\,kpc derived using Ca\,\II~lines \citep{2009A&A...507..833M}. Previously, \textit{Gaia} Data Release~2 (DR2) \citep{2018yCat.1345....0G} provided a distance of 2.66$\pm$0.23\,kpc \citep{2018AJ....156...58B}. Considering that \textit{Gaia} DR3 provides a distance that is concordant with other measurements, we adopt 2.28$\pm$0.09\,kpc to be the most accurate distance value.

We obtain \acp{PM} from \textit{Gaia} DR3 \citep{2020yCat.1350....0G}, for WR16, which are $\mu_{\alpha}$:~$-9.458\pm 0.021$\,mas\,yr$^{-1}$ and $\mu_{\delta}$:~$5.054\pm 0.018$\,mas\,yr$^{-1}$. Using the Gaia parallax and proper motion values, we calculate the peculiar velocity with respect to the local interstellar medium following \citet{2007A&A...467L..23C} and \citet{2020MNRAS.495..417C}. We determine the stellar peculiar velocity to be  $V_{\alpha}(pec) =$ --45.3$\pm$5.4\,\kms\ and $V_{\delta}(pec) =$ 22.8$\pm$4.7\,\kms, indicating that the star is moving in a north-west direction, which is discussed further in section \ref{Sub_Section:Shell_Expansion}. This matches the brighter north-western component of the circular nebulosity observed in radio, infrared, and H$\alpha$ (Figures \ref{fig:WR16/WR40} and \ref{fig:IR-Ha}). We then calculate the peculiar tangential velocity to be 50.7$\pm$6.9\,\kms.

\begin{figure}[ht]
\centering
    \includegraphics[width=\linewidth]{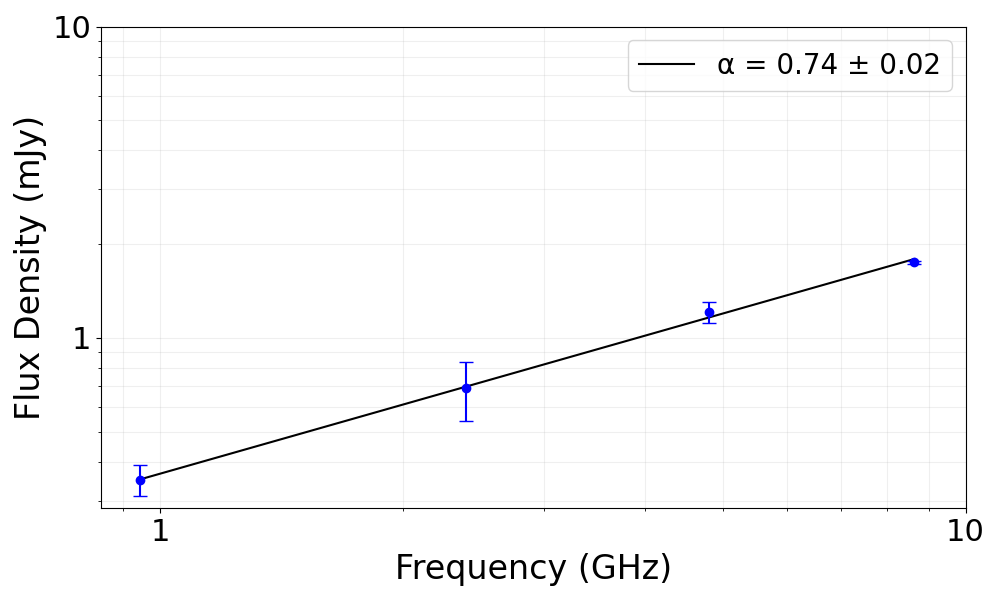}
    \caption{Spectral index plot of WR16 star, using flux density values from Table~\ref{Table:Data_Fluxes}.}
    \label{fig:specindex}
\end{figure}



We measure the inner WR16 nebula to have a flux density of 72.2$\pm$7.2\,mJy at 943.5\,MHz, taking an error of 10\% \citep[as discussed in ][]{2024MNRAS.534.2918S,2024PASA...41..112F}. We also measure the flux density of the central star at 943.5\,MHz, and find it to be 0.35$\pm$0.04\,mJy. We combine this with the \ac{ATCA} flux density measurements listed in Table~\ref{Table:Data_Fluxes} and section \ref{Sub_Section:Data_Other} to compute a spectral index for the star. The radio spectral index is defined as $S \propto \nu^{\alpha}$, where $S$ is flux density, $\nu$ is the frequency and $\alpha$ is the spectral index. We calculate the spectral index using the \textsc{linregress}\footnote{\url{https://docs.scipy.org/doc/scipy/reference/generated/scipy.stats.linregress.html}} function in the Python \textsc{scipy}~\citep{Virtanen2020} library. The function uses the linear least-squares regression method to find a best fit line of $
\alpha\,=\,+0.74\pm0.02$ for WR16. This is very close to the canonical spectral index of an isothermal, spherical stellar wind ($\alpha=+0.6$, \citep{1975A&A....39....1P, 1975MNRAS.170...41W}), and is similar to that of \citet{2000MNRAS.319.1005D} for similar \ac{WR} stars that also exhibit thermal characteristics.




We fit a circular region to the outer edge of the shell and measure an angular diameter of 8\farcm42, which is similar to the angular size estimated in \citet{2020MNRAS.495..417C}. We use the derived distance and the angular size measured on the \ac{EMU} image (Figure~\ref{fig:WR16/WR40}) to determine the shell's true linear size and find that the shell has a diameter of 5.57$\pm$0.22\,pc. Due to the shell's presence in near--infrared and H$\alpha$ (Figure \ref{fig:IR-Ha}) we infer that the shell also has thermal origin. Because the shell has not yet been observed at other frequencies, we cannot confirm thermal origin from the spectral index.



\subsection{Large-Scale Wind-Blown Bubble of CO}
\label{Sub_Section:CO_Map_Discussion}

Figure~\ref{Figure:CO_Map}(a) shows the velocity-integrated intensity map of CO toward WR16 with an integration range of from $-7$ to 0~km~s$^{-1}$. We found a new candidate for a large-scale wind-blown bubble whose radius is approximately 0\fdg75 or $\sim$30 pc at the distance of WR16. Since the inner circular shell of WR16 is approximately placed at the geometric center of the CO bubble, they appear to be physically associated to each other.

Figures~\ref{Figure:CO_Map}(b) and \ref{Figure:CO_Map}(c) show the position--velocity diagrams of CO. We found clear evidence of an expanding gaseous bubble whose expanding velocity is $\sim$7~km~s$^{-1}$. The fact that the spatial extent in the Declination direction on the position--velocity diagram varies depending on the integrated Right Ascension range is also roughly consistent with the three-dimensional expansion motion of the molecular clouds.

We argue that the dynamical timescale of this expanding gaseous bubble can be naturally understood if it is assumed to have been driven by WR16. The dynamical time scale of the expanding bubble is estimated to be (bubble radius)/(expanding velocity) $\sim$ 30~pc / 7~km~s$^{-1}$ $\sim$ 4~Myr. This time scale is roughly consistent with the typical lifetime of a massive star, suggesting that the CO expanding bubble was likely formed by stellar feedback effects such as strong stellar winds from the progenitor of WR16.

\subsection{Shell Expansion}
\label{Sub_Section:Shell_Expansion}

\begin{figure*}[!ht]
\centering
    \includegraphics[width=\linewidth]{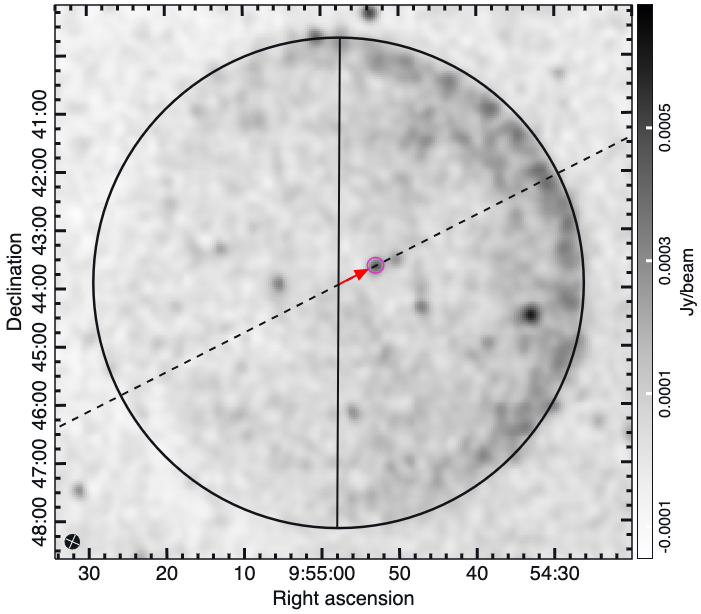}
    \caption{Measurements of the circular nebulosity surrounding WR16 and peculiar velocity mapped over the EMU image, similar to Figure \ref{fig:WR16/WR40}. The 8\farcm42 diameter shell is shown as a black circle. The magenta circle indicates the position of WR16, and the red arrow shows its distance from the shell's geometric center, as well as its direction. The black dashed line represents the projected path of WR16’s peculiar velocity, which points toward the north-west (top-right).}
    \label{Figure:WR16_Measurements}
\end{figure*}

Figure~\ref{Figure:WR16_Measurements} shows several on-image measurements of the shell surrounding WR16. We find that the tangential peculiar velocity of the star passes through the geometric center of the 8\farcm42 circle (RA (J2000) = 09:54:57.8, Dec (J2000) = --57:43:58.9), as indicated by the dashed black line in the figure. The distance of the star to the center of the nebulosity is 44\farcs81. We convert this distance from angular size to linear size (0.49~pc, or $1.53\times 10^{13}$\,km), and multiply the calculated velocity in Section \ref{Sub_Section:Measurements} by seconds in a year ($3.154\times 10^{7}$\,s) to obtain a yearly velocity of $1.58\times 10^{9}$~km\,yr$^{-1}$, $5.12\times 10^{-5}$~pc~yr$^{-1}$. We can determine the amount of time it has taken for WR16 to travel to its current point by dividing the linear size by the yearly velocity, and return a value of $\sim 9500\pm1300$\,yr.

Assuming that the shell originated at the point that WR16 passed through its geometric centre coordinates, we can measure its average expansion velocity based on the distance travelled by WR16. The radius of the shell is 4\farcm21, which converts to a linear size of $8.62\times 10^{13}$\,km (2.79~pc). Dividing this by the years travelled ($\sim 9500\pm 1300$\,yr) and dividing by seconds in a year, we determine the average expansion velocity of the circular shell to be $280\pm 40$\,\kms. It is important to note that this only assumes a 2--dimensional plane, and is therefore a lower limit on age values and an upper limit on expansion values.


We take our calculated expansion velocity to be the average rate of expansion because it is presumed that the shell is slowing down the more it expands and interacts with the surrounding interstellar medium. WR16 also has a wind terminal velocity ($v_{\infty}$) of 630~km~s$^{-1}$~\citep{2015A&A...578A..66T}, so it is likely that the wind-driven shell started with velocities similar to this, and has slowed over the $\sim 9500\pm 1300$\,yr timescale. This velocity sits conveniently between the typical expansion speed of \ac{LBV} shells of $\sim$50 km~s$^{-1}$ \citep{2011IAUS..272..372W}, like that of AG Car \citep{1991IAUS..143..385S}, and the faster ejecta of Eta Car \citep[600~km~s$^{-1}$,][]{2014MNRAS.442.3316S}. This supports the idea that the shell originated in a previous \ac{LBV} phase. 



\subsection{Mass-Loss Rate}
\label{Sub_Section:Mass_Loss}

Based on the observed parameters of WR16 at 943.5~MHz, we are able to determine a mass-loss rate ($\dot{M}$) at this frequency using the equations outlined in \citet[][Equation 20]{1975MNRAS.170...41W} and \citet[][Equation 3]{1995ApJ...450..289L}. We calculate a Gaunt factor, $g_v$, at 943.5~MHz to be 5.99, assuming a gas temperature of 10$^4$~K and a $Z$ of 1. Using the same constants for our calculation as \citet{1995ApJ...450..289L}, and substituting our flux, distance, and frequency, we derive a mass-loss rate of $1.753\times 10^{-5}~M_\odot~$yr$^{-1}$. 

This is lower than the calculated mass-loss rate from \citet{1995ApJ...450..289L} of $3.981\times 10^{-5}~M_\odot~$yr$^{-1}$. However, recalculating their value using our \textit{Gaia} distance measurement of 2.28~kpc, we get a result of $2.188\times 10^{-5}~M_\odot~$yr$^{-1}$. These measurements are quite similar, and may indicate that the mass-loss rate has decreased slightly over the 30~yrs between observations. It is also important to note that the canonical value for mass-loss rates in WN8 stars like WR16 (Section \ref{Section:Introduction}) is $1.995\times 10^{-5}~M_\odot~$yr$^{-1}$ \citep{2007ARA&A..45..177C}.

Using the calculated mass-loss rate, we are also able to determine a lower limit of ionising photons that WR16 are released per second as in \citet[][Equation 23]{1975MNRAS.170...41W}. We use a radius of 19~$R_\odot$ \citep{1989A&A...210..236S} to calculate a lower limit value of $N_{UV} > 1.406\times 10^{47}~$s$^{-1}$, which is in concordance with the canonical value for WN8 stars of $1.259\times 10^{49}~$s$^{-1}$ \citep{2007ARA&A..45..177C}.


\subsection{Similarities to Other Objects}
\label{Sub_Section:Similarities}

WR16 presents an almost symmetrical circular shell similar to the Galactic \ac{SNR} Teleios (Greek $\tau\epsilon\lambda\epsilon\iota o\varsigma$ -- meaning perfect, Filipovi{\' c} et al. submitted). Although they share similarities, they are unique objects for their given type. This shows that different objects can share the same morphologies, and can possibly be equated to similar formation processes.

Another object similar to the WR16 shell is the radio ring K\'yklos, discussed in \citet{2024A&A...690A..53B}.They discuss the likelihood of the ring-like object being an outburst of a \ac{WR} star. This object is also situated in our Galaxy, but is significantly smaller than WR16 (WR16: 8\farcm42, K\'yklos: 80$^{\prime\prime}$). Both systems show evidence of thermal emission, however, K\'yklos presents a ring-like structure rather than a diffuse circle like WR16. The ring-like structure and smaller size may indicate that K\'yklos is at an earlier developmental stage compared to WR16.


We see significant brightening toward the north--west part of the WR16 circular shell. This is likely due to the closer proximity of WR16, which also explains the much fainter south--east portion. This edge brightening is similar to the Lagotis \HII\,region discussed in \citet{2025PASA...42...32B}. In the case of Lagotis, edge brightening is caused by a cluster moving into a molecular cloud, whereas for WR16, it can be seen as the star catching up with its shell edge.


\section{Conclusion}
\label{Section:Conclusion}

We present the detection of the nebulosity surrounding the Wolf-Rayet star WR16 and its circular shell at 943.5\,MHz using the ASKAP EMU survey. We find that the shell has a measured flux density of 72.2$\pm$7.2\,mJy, and WR16 has a flux density of 0.35$\pm$0.04\,mJy. By combining this with archival \ac{ATCA} data, we determine a spectral index of $+$0.74$\pm$0.02 for the star, which indicates thermal origin of the emission. We also infer that the shell is thermal from its presence in IR and H$\alpha$. We use \textit{Gaia} DR3 measurements to determine the star's direction and distance. WR16 is moving in a north-western direction, coinciding with the brighter edge of the nebulosity. 

We also find that the star is at a distance of 2.28$\pm$0.09\,kpc, and measure the shell to be 8\farcm42 in angular diameter, which translates to a linear size of 5.57$\pm$0.22~pc. The shell as seen by \ac{EMU} matches well with \ac{WISE} and SuperCOSMOS data, further confirming thermal emission. By mapping \ac{PM}, we are able to determine the origin point of the shell and calculate the age of the shell to be $\sim 9500\pm 1300$\,yr, with an average expansion velocity of $280\pm 40$~km~s$^{-1}$. It is expected that the brighter edge of the shell will become brighter as WR16 moves towards it. 

We also discuss the discovery of a large wind-blown bubble seen in CO observations, which is proposed to be related to WR16 during its main-sequence phase. We calculate a mass-loss rate of $1.753\times 10^{-5}~M_\odot~$yr$^{-1}$, which is remarkably close to the canonical value of the mass-loss rate for WN8 type stars. Using this we were able to determine the amount of UV-ionising photons emanating from WR16: $N_{UV} > 1.406\times 10^{47}~s^{-1}$. 


\paragraph{Acknowledgements}

This scientific work uses data obtained from Inyarrimanha Ilgari Bundara, the CSIRO Murchison Radio-astronomy Observatory. We acknowledge the Wajarri Yamaji People as the Traditional Owners and native title holders of the Observatory site. CSIRO’s ASKAP radio telescope is part of the Australia Telescope National Facility (\url{https://ror.org/05qajvd42}). Operation of ASKAP is funded by the Australian Government with support from the National Collaborative Research Infrastructure Strategy. ASKAP uses the resources of the Pawsey Supercomputing Research Centre. Establishment of ASKAP, Inyarrimanha Ilgari Bundara, the CSIRO Murchison Radio-astronomy Observatory and the Pawsey Supercomputing Research Centre are initiatives of the Australian Government, with support from the Government of Western Australia and the Science and Industry Endowment Fund. 

This work has made use of data from the European Space Agency (ESA) mission {\it Gaia} (\url{https://www.cosmos.esa.int/gaia}), processed by the {\it Gaia} Data Processing and Analysis Consortium (DPAC, \url{https://www.cosmos.esa.int/web/gaia/dpac/consortium}). Funding for the DPAC has been provided by national institutions, in particular the institutions participating in the {\it Gaia} Multilateral Agreement.

This work was also supported by JSPS KAKENHI grant No. 21H01136 (HS), 24H00246 (HS). The NANTEN project is based on a mutual agreement between Nagoya University and the Carnegie Institution of Washington (CIW). YF greatly appreciate the hospitality of all the staff members of the Las Campanas Observatory of CIW. YF and HS are thankful to many Japanese public donors and companies who contributed to the realisation of the project.

MDF, SL and GR acknowledge Australian Research Council (ARC) funding through grant~DP200100784.

CJR acknowledges financial support from the German Science Foundation DFG, via the Collaborative Research Center SFB1491 `Cosmic Interacting Matters – From Source to Signal'.

\paragraph{Data Availability Statement}

\ac{EMU} data can be accessed through the \ac{CASDA} portal: \url{https://research.csiro.au/casda}. All \textit{Gaia} \ac{DR3} data are obtained from the \textit{Gaia} Archive website: \url{https://gea.esac.esa.int/archive/}. \ac{WISE} data are available from the NASA/IPAC Infrared Science Archive (IRSA): \url{https://irsa.ipac.caltech.edu/Missions/wise.html}.
The SuperCOSMOS H$\alpha$ data are available from The Wide Field Astronomy Unit (WFAU) archive: \url{http://www-wfau.roe.ac.uk/sss/halpha/hapixel.html}

\printendnotes

\bibliography{WR16}

\end{document}